\documentclass[conference,10pt]{IEEEtran}
\IEEEoverridecommandlockouts
\usepackage{cite}
\usepackage{amsmath,amssymb,amsfonts,mathrsfs}
\usepackage{algorithmic}
\usepackage{graphicx}
\usepackage{textcomp}
\usepackage{braket}
\def\BibTeX{{\rm B\kern-.05em{\sc i\kern-.025em b}\kern-.08em
    T\kern-.1667em\lower.7ex\hbox{E}\kern-.125emX}}

\UseRawInputEncoding

\begin{document}

\title{Canonical Distribution of the Occupancy Numbers of Bosonic
Systems
\\
}

\author{Arnaldo Spalvieri \\
Dipartimento di Elettronica, Informazione e Bioingegneria \\
Politecnico di Milano\\
arnaldo.spalvieri@polimi.it \\
ORCID 0000-0002-8336-7996
}
\maketitle
\begin{abstract}

The paper works out the canonical probability distribution of the
occupancy numbers of a bosonic system and shows that canonical
typicality applies to the canonical density operator of the
occupancy numbers. The result is that, if, as it is today
standard, the canonical system's mixed state is obtained by
tracing out the environment from any typical pure state of the
universe, then asymptotically the canonical probability
distribution of system's occupancy numbers tends in probability to
the multinomial distribution. The paper also shows that the
currently accepted probability distribution of the occupancy
numbers of a system with fixed number of particles is not
compatible with the commonly accepted notion of canonical system.

\vspace{0.5cm}

{\em Keywords:}\hspace{0.2cm}{\bf Canonical Systems, Occupancy
Numbers, Canonical Typicality, Multinomial Distribution.}

\end{abstract}

\section{Introduction}

More than a century ago, Bose and Einstein derived the probability
distribution of the occupancy numbers of the photon gas in
\cite{bose} and of the gas of massive particles in
\cite{einstein24}. The Bose-Einstein approach applies to a system
of the grand-canonical ensemble, that is a system that
absorbs/emits particles, e.g. photons in Bose's analysis of the
black-body radiation, from/to the surrounding environment. Due to
the random absorption/emission of particles, the total number of
system's particles randomly evolves in time. In contrast, a closed
system, for instance a gas of massive particles trapped inside a
region of space, is made by a fixed number of particles. For this
reason, the probability distribution of the occupancy numbers of a
closed bosonic system cannot be the Bose-Einstein product of
geometric distributions.

Actually, it is widely recognized that the Bose-Einstein analysis
of a closed system near the condensation point leads to the
so-called {\em grand-canonical catastrophe}, as it predicts
fluctuations of the ground-state occupation that are
unrealistically large, see for instance \cite{kocha}. A consistent
analysis therefore requires a treatment in which the number of
particles is fixed. However, this introduces a difficulty, because
the constraint on the number of particles induces dependencies
between the random occupancy numbers, dependencies that instead
are not present in the original Bose-Einstein approach.

The currently accepted approach, proposed by Ziff, Uhlenbeck, and
Kac in \cite{ziff}, works out the partition function of a closed
system of $N$ particles by taking the $N$-th coefficient of the
grand-canonical generating function. The extensive bibliography of
\cite{barghathi} bears witness to the effort made by the
scientific community to advance the understanding of the
properties of the distribution proposed in \cite{ziff}.

We intentionally don't go through the long history of this really
vast research line (for this, the reader is referred to
\cite{kocha}) because our main point is that, as we show in the
paper, the distribution proposed in \cite{ziff}, although
referring to a system with fixed number of particles and although
leading to an accurate analysis of the condensation phenomenon, is
not compatible with the commonly accepted notion of canonical
system. This observation motivates us to re-consider the canonical
probability distribution of the occupancy numbers of a bosonic
system. In doing this, we start from a suggestion sketched in 1.6
of the book by Pathria and Beale \cite{pathria}, a suggestion
that, however, is not carried through to the full derivation of
the desired probability distribution neither in \cite{pathria} nor
in other papers we are aware of.

A quick overview of the paper is the following. In the second
section we give the definitions. In the third section we derive
the canonical density operator of the occupancy numbers of a
bosonic system. This is done by mapping the standard canonical
density operator living in system's full Hilbert space onto the
symmetric Hilbert subspace. What we find is that that the
probability distribution that equips the sought density operator
is multinomial. In the fourth section we prove that, when the
system is obtained by tracing out the environment from the
universe (the union of system and of the very much larger
environment that surrounds the system), the vast majority of the
pure states of the universe actually lead to a density operator of
the occupancy numbers close to the canonical one, or, in other
words, that the canonical typicality principle of
\cite{typpopescu,typgold} applies also to the canonical density
operator of the occupancy numbers. Upper bounds in the style of
those obtained in \cite{typpopescu} for systems living in the full
Hilbert space are obtained here for the bosonic system in the
symmetric Hilbert subspace. In the fifth section we consider the
special case in which the categorical distribution that equips the
multinomial distribution is the Boltzmann distribution. In this
section, we also show that the currently accepted probability
distribution of \cite{ziff} does not match a Boltzmannian
canonical system, especially when it has a small number of
particles. Putting ourselves in the perspective of \cite{ziff}, we
discuss the reasons behind this mismatch and why our proposed
approach actually overcomes the issue.

\section{Occupancy numbers of a system with fixed and
known number of bosons}

Consider a system of $N$ non-relativistic particles of the same
species, let ${\cal C}=\{1,2, \cdots,|{\cal C}|\}$, be the set of
quantum eigenstates allowed to a particle, let $c_i \in {\cal C},
\ i=1,2, \cdots, N,$ represent the generic eigenstate of the
$i$-th particle, and let ${s}=(c_1,c_2,\cdots,c_N)$, ${s} \in
{\cal S}={\cal C}^N$, represent the generic eigenstate of the
system (the so-called {\em microstate} in many textbooks, e.g.
\cite{pathria}). The number of particles that occupy the
one-particle eigenstate $c$ is
\begin{align} n_c({s})=\sum_{i=1}^N \delta_{c,c_i},  \nonumber \end{align} where
\begin{align} \delta_{x,y}
=\left\{
\begin{array}{cc}  1, &
 \mbox{if} \ x=y,\\
0,  &  \mbox{elsewhere}.
\end{array} \right.
\nonumber\end{align} In the following, when unnecessary we will
omit the dependance of the occupancy number on ${s}$ and we will
denote $n$,
\[{n}=(n_1,n_2, \cdots, n_{|{\cal C}|}),\]
the vector of the occupancy numbers, or, in short, the occupancy
vector, being understood that $n$ must satisfy the constraint
\[\sum_{c}n_c=N.\]  The size $|\cal{N}|$ of the set $\cal{N}$ of the occupancy
vectors allowed to a system with $N$ particles is
\[|{\cal N}|=\frac{(N+|{\cal C}|-1)!}{N!(|{\cal C}|-1)!}.\]

Let the subset ${\cal S}_{{n}}$ of ${\cal S}$ be the set of all
the vectors ${{s}}$ that have the same vector ${n}({{s}})$ of
occupancy numbers. ${\cal S}_{{n}}$ contains only and all the
$W_{{n}}$ permutations of anyone of its vectors, where
 \begin{align} W_{{n}} =\left\{
\begin{array}{cc}  \frac{N!}{\prod_c
n_c!}, &
 \mbox{if} \ \sum_{c}n_c=N,\\
0,  &  \mbox{elsewhere},
\end{array} \right.
\nonumber \end{align} is the multinomial coefficient. When ${n}$
spans the entire set ${\cal N}$, the set $\{{\cal S}_{{n}}\}$ form
a disjoint partition of ${\cal S}$,
\[\bigcup_{{n}\in {\cal N}}{\cal
S}_{{n}}={\cal S}, \ \ \bigcap_{{n}\in {\cal N}}{\cal
S}_{{n}}=\emptyset,\] hence each $s \in {\cal S}$ belongs to one
and only one of the elements of $\{{\cal S}_n\}$.

\section{Canonical density operator in the symmetric Hilbert subspace}

Let the universe consist of $U$ of bosons, denote its Hilbert
space ${\cal H}_{\cal U}={\cal H}_{\cal C}^{\otimes U}$, where
$\otimes$ denotes the Kronecker product and ${\cal H}_{\cal C}$ is
the Hilbert space of the quantum states allowed to one boson, and
consider the restriction of the Hilbert space ${\cal H}_{{\cal
U},\epsilon}\subseteq{\cal H}_{\cal U}$ spanned by the eigenstates
belonging to the strongly typical set ${\cal T}_{{\cal
U},\epsilon}$,
\begin{align} {\cal T}_{{\cal U},\epsilon}=  \left\{\tau:
\left|\frac{n_c(\tau)}{U}-P_c\right|< \epsilon, \ \forall c \in
{\cal C} \right\},\label{typset}
\end{align}
where $\epsilon>0$ is a small number,
\[\tau=(c_1,c_2, \cdots, c_U),\]
and $P_c$ is the limit at which, as $U \rightarrow \infty$, tends
the empirical relative frequency whichever is $\tau$ in the
strongly typical set. We define $\rho_{\epsilon}$ as the density
operator of the reduced state of the system of $N$ particles
obtained from the maximally mixed state in the restriction of the
universe:
\begin{align} \rho_{\epsilon}=  \mbox{Tr}_E (\rho_{{\cal U},\epsilon,\max}) ,\label{canprob1}
\end{align}
where $\mbox{Tr}_E(\cdot)$ traces out the environment from the
universe and
\begin{align} \rho_{{\cal U},\epsilon, \max}=  \frac{1}{|{\cal T}_{{\cal
U},\epsilon}|}\sum_{\tau \in {\cal T}_{{\cal U},\epsilon}}
\ket{\tau}\bra{\tau}\label{maximallymixed}
\end{align}
is the density operator of the maximally mixed state in the
restriction of the universe.

The  density operator $\rho_{{\cal U},\epsilon, \max}$ is
invariant to permutations of the single-particle eigenstates
$\ket{c_1},\ket{c_2},\cdots,\ket{c_U},$ because all the
eigenstates of the same type (i.e., with the same occupancy
numbers) contribute to it with the same weight. A consequence of
the invariance to permutations is that, as $U \rightarrow \infty$,
by the quantum de Finetti theorem \cite{caves,konig}, any
$N$-body, $N\leq U$, reduced state obtained from $\rho_{{\cal
U},\epsilon, \max}$ tends in probability to a mixture of
independent and identically distributed (i.i.d.) product states:
\begin{align} \rho_{\epsilon} &\xrightarrow[]{\text{in probability}}
\left(\sum_cP_c\ket{c}\bra{c}\right)^{\otimes N} \label{definetti}
\\ &=\sum_sP_c^{n_c(s)}\ket{s}\bra{s}=\rho_{can},\label{canprob2}
\end{align}
where the last equality means that we take its left hand side as
the definition of system's canonical density operator
$\rho_{can}$. The i.i.d. property means that the probability
distribution of microstates $\{P_s\}$ on which the diagonal
density operator $\rho_{can}$ is based, is equal to the product of
identical factors:
\begin{align} P_{s} =
\prod_{i=1}^NP_{c_i}=\prod_{c}P_c^{n_c(s)}. \nonumber
\end{align}

When the system consists of indistinguishable bosons, the
symmetrization postulate requires that the physical Hilbert space
is the symmetric subspace of the full Hilbert space. System's
Hamiltonian, being invariant under particle permutations,
preserves this symmetric subspace. The symmetric Hilbert subspace
${\cal H}_{\cal N}$ of system's Hilbert space ${\cal H}_{\cal
S}={\cal H}_{\cal C}^{\otimes N}$, is spanned by the complete set
of ''bosonic'' eigenstates $\{\ket{{n}}, \ {n} \in {\cal N}\},$
with
\begin{align}\ket{{n}}= \sum_{{s} \in {\cal
S}_{{n}}}\frac{\ket{{s}}}{\sqrt{W_{{n}}}}, \nonumber 
\end{align}
see \cite{church} for a tutorial on the topic.

The following family of Kraus operators operates the mapping from
${\cal H}_{\cal S}$ to ${\cal H}_{\cal N}$:
\begin{align} K_{{n},{s}}
=\left\{
\begin{array}{cc} \ket{{n}}\bra{{s}} , &
 \mbox{if} \ {s} \in {\cal S}_{n},\\
0,  &  \mbox{elsewhere}.
\end{array} \right.
\nonumber 
\end{align}
The mapping is complete because
\begin{align} \sum_{{n}} \sum_{{s} }K_{{n},{s}}^{\dagger}K_{{n},{s}} &=
\sum_{{s}}\ket{s}\bra{s}, \nonumber
\end{align}
therefore the above family of Kraus operators defines a CPTP
linear map from ${\cal H}_{\cal S}$ to ${\cal H}_{\cal N}$.  We
obtain the ''bosonic'' canonical density operator ${\nu}_{can}$ by
mapping ${\rho}_{can}$ onto the symmetric subspace:
\begin{align}{\nu}_{can}&=
\sum_{{n}} \sum_{{s}}K_{{n},{s}}
\rho_{can}K_{{n},{s}}^{\dagger}\nonumber
\\ &=
\sum_{{n} } \sum_{{s} \in {\cal S}_{{n}}}\ket{{n}}\braket{{s}|
\rho_{can}|{s}}\bra{{n}}\nonumber
\\ &=\sum_{{n} }\sum_{{s} \in {\cal S}_{{n}}}P_c^{n_c}\ket{{n}}\bra{{n}}
\nonumber
\\ &=\sum_{{n} }P_n \ket{{n}}\bra{{n}},
\nonumber 
\end{align}
with
\begin{align}P_n= W_{{n}}\prod_{c}
P_c^{n_c}.
 \label{multinomial}
\end{align}

The above equality  shows that the canonical probability
distribution of the occupancy numbers is the multinomial
distribution, see also \cite{pnas,zupa} for the multinomial
distribution in statistical mechanics. In the context of
probability theory, the one-particle probability distribution
$\{P_c\}$ that equips the multinomial distribution is called the
{\em categorical} distribution, the $|{\cal C}|$ categories being
here the $|{\cal C}|$ eigenstates.

We observe that the product $\prod_{c}n_c!$ contained in $W_n$ is
a factor of the denominator of (\ref{multinomial}). Due to its
presence, the canonical distribution (\ref{multinomial}) of the
occupancy numbers is not simply proportional to the product of
factors, hence, as desired, the occupancy numbers are not
independent random variables, while, as already mentioned, the
probability distribution of microstates has the i.i.d. form.

\section{Bosonic canonical typicality}

In \cite{spalvmacro} we obtained ${\nu}_{can}$ by tracing out the
environment from the universe in the case in which, according to
the eigenstate thermalization hypothesis of \cite{eth}, the state
of the universe is the ''thermal'' bosonic eigenstate. In other
words, our previous result refers to the case in which the set
${\cal T}_{{\cal U},\epsilon}$ contains only and all the
eigenstates with the same occupation numbers, or, in the language
of typicality, only one type. We now substantially broaden the
scope of our previous result by tracing out the environment from
any pure state picked with uniform probability from ${\cal
H}_{{\cal U},\epsilon}$.

Let the universe be in any pure state $\ket{\phi}$ and consider
system's density operator
\[\rho_{\phi}=\mbox{Tr}_E(\ket{\phi}\bra{\phi}).\]
Using Levy's lemma,
 paper \cite{typpopescu} proves that, for $\ket{\phi}$ randomly
 picked with uniform probability inside the restricted Hilbert space
 ${\cal H}_{{\cal U},\epsilon}$ of the universe and for any $\eta>0$,
\begin{align}
\hspace{-0.2cm}\Pr(\,\| \rho_{\phi}-\rho_{\epsilon}\|_1 \ge \eta
\,) &\le 2\exp\!(-C\,|{\cal T}_{{\cal U},\epsilon}|([\eta -
\mu_{\Delta \rho}]_+)^2),\label{PSW2}
\end{align}
where
\begin{align}|| \rho||_1 = \mbox{Tr}(\sqrt{\rho
\rho^{\dagger}}) \nonumber
\end{align}
is the trace norm of the operator $\rho$, $C$ is a constant coming
from Levy's lemma,
\[[\eta - \mu_{\Delta \rho}]_+=\max\{\eta - \mu_{\Delta
\rho},0\},\] and
\[\mu_{\Delta \rho}=E\{\| \rho_{\phi}-\rho_{\epsilon}\|_1\},\]
where $E\{\cdot\}$ is the expectation over the random variable
inside the curly brackets, in this case, the random $\phi$. The
nontrivial exponential decay is obtained for
\[\eta \geq \mu_{\Delta \rho}.\]
A conservative estimate of the value of $\eta$ at which the decay
starts to be exponential is the left hand side of
\begin{equation}\sqrt{|{\cal S}| / d_E^{\rm eff}}\geq \mu_{\Delta
\rho},\nonumber \end{equation} where $d_E^{\rm
eff}=1/\mbox{Tr}((\mbox{Tr}_S(\rho_{{\cal U},\epsilon,\max}))^2)$
is the effective dimension of the environment and the inequality
is the main result of \cite{typpopescu}. For $U \gg N$, $d_E^{\rm
eff}$ is much greater than $|{\cal S}|$ and values of $\eta$ much
greater than the very small number $\sqrt{|{\cal S}| / d_E^{\rm
eff}}$ are of interest. In this case, the bias term $\mu_{\Delta
\rho}$ inside the exponent of (\ref{PSW2}) can safely be
neglected.

Now let \(|{\cal N}_{{\cal U},\epsilon}|\) denote the number of
distinct occupation vectors (types) inside the restriction of the
universe.
 Since all microstates
of the same type are mapped onto the same occupancy vector, the
trace norm of the difference between the two density operators in
system's symmetric subspace depends only on the $|{\cal N}_{{\cal
U},\epsilon}|$ occupation vectors. Therefore, applying Levy's
lemma with dimension $|{\cal N}_{{\cal U},\epsilon}|$ equal to the
dimension of the restricted symmetric subspace of the universe,
the tail bound becomes
\begin{equation}\label{eq:refined-bound}
\hspace{-0.0cm}\Pr(\| \nu_{\phi}-\nu_{\epsilon}\|_1 \ge \eta) \le
2\exp(-C|{\cal N}_{{\cal U},\epsilon}|([\eta-\mu_{\Delta
\nu}]_+)^2).\hspace{-0.05cm}
\end{equation}
By the same arguments of \cite{typpopescu} it can be proved that
the upper bound above the bias becomes
\[\sqrt{|{\cal N}|/ d_{E,sym}^{\rm eff}}\geq \mu_{\Delta \nu},\]
where $d_{E,sym}^{\rm eff}=1/\mbox{Tr}((\mbox{Tr}_S(\nu_{{\cal
U},\epsilon,\max}))^2)$ is the effective dimension of the
environment in the symmetric subspace.

By typicality arguments it can be shown that
\begin{equation}|{\cal T}_{{\cal U},\epsilon}|\sim 2^{-U \sum_cP_c
\log_2(P_c)}.\label{typsize}
\end{equation}
At the same time,
\[|{\cal N}_{{\cal U},\epsilon}|\leq
\frac{(U+|{\cal C}|-1)!}{U!(|{\cal C}|-1)!}\]  scales only
polynomially with the number of particles of the universe, not
exponentially. Therefore, in the thermodynamic regime of the
universe, in which \(|{\cal N}_{{\cal U},\epsilon}|\ll |{\cal
T}_{{\cal U},\epsilon}|,\) and for values of $\eta$ of practical
interest, the bound \eqref{eq:refined-bound} can be much tighter
than the original bound (\ref{PSW2}), which, by contractivity of
the CPTP map, remains valid in the symmetric subspace as a
conservative estimate.

\section{Boltzmann's categorical distribution}

Let us consider the Boltzmann one-particle probability
distribution:
\begin{align} P_c= \frac{e^{-\beta \epsilon_c}}{Z},
\ \  Z=\sum_{c} e^{-\beta \epsilon_c}, \label{boltzmann}
\end{align}
where $\epsilon_c$ is the energy eigenvalue of eigenstate $c$ and
$\beta>0$ is a real scalar. Taking (\ref{boltzmann}) as the
categorical distribution of the multinomial distribution, for the
probability distribution of microstates and for the probability
distribution of the occupancy numbers we get
\begin{align} P_s&= \frac{\prod_{i=1}^Ne^{-\beta \epsilon_{c_i}}}
{Z^N}= \frac{\prod_{c} e^{-\beta \epsilon_c
n_c(s)}}{\sum_{n}W_{{{n}}}\prod_{c} e^{-\beta n_c \epsilon_c}},
\label{boltzmannmany}
\end{align}
\begin{align} P_n&=  \frac{W_n\prod_{c} e^{-\beta \epsilon_c
n_c}}{\sum_{{n}}W_{{{n}}}\prod_{c} e^{-\beta n_c \epsilon_c}},
\nonumber 
\end{align}
where we have substituted the multinomial expansion of $Z^N$,
\begin{align} Z^N=\left(\sum_{c } e^{-\beta
\epsilon_c}\right)^N=\sum_{{n} }W_{{{n}}}\prod_{c} e^{-\beta n_c
\epsilon_c}. \nonumber 
\end{align}

For completeness, hereafter we show that the probability
distribution of occupancy numbers proposed in eqn. (1.127) of
\cite{ziff} for a system at the thermal state, which is
\begin{align}P_{n,\small{\text{Ziff}}}&=\frac{\prod_{c}
e^{-\beta \epsilon_c n_c}}{Z_{\small{\text{Ziff}}}},\label{ziff}
\end{align}
with
\begin{align}Z_{\small{\text{Ziff}}}&=\sum_{n }\prod_{c} e^{-\beta \epsilon_c n_c}
\label{gottlieb} \\&  =\sum_{n}W_{n}^{-1}\sum_{s \in {\cal
S}_n}\prod_{c} e^{-\beta \epsilon_c n_c(s)}\nonumber \\
&=\sum_{s}W_{n(s)}^{-1}\prod_{c} e^{-\beta \epsilon_c n_c(s)},
\label{borrmann}
\end{align}
cannot be an attribute of a Boltzmannian canonical system. We
write the partition function in the two equivalent forms
(\ref{gottlieb}) and (\ref{borrmann}) just to explicitly identify
(\ref{gottlieb}) with eqn. (1.124) of\cite{ziff}, and
(\ref{borrmann}) with eqn. (5) of \cite{borrmann}.

We observe that, if (\ref{boltzmannmany}) holds, then the mutually
exclusive $W_n$ microstates belonging to ${\cal S}_n$ are equally
probable. Assuming equiprobability of the $W_n$ microstates
belonging to ${\cal S}_n$, the probability of a microstate $s$
with occupancy numbers $n$ obtained from (\ref{ziff}) is
\begin{equation}\frac{P_{n(s),\small{\text{Ziff}}}}{W_{n(s)}}=\frac{\prod_{c}
e^{-\beta \epsilon_c
n_c(s)}}{W_{n(s)}Z_{\small{\text{Ziff}}}}.\label{reif2}\end{equation}
Since $W_{n(s)}Z_{\small{\text{Ziff}}}\neq Z^N$, we see that
(\ref{reif2}) is not compatible with (\ref{boltzmannmany}), which
not only descends from (\ref{definetti}), but also is how is
commonly understood a canonical Boltzmannian system.

Incompatibility between (\ref{reif2}) and (\ref{boltzmannmany})
arises because the construction of \cite{ziff} is derived from the
grand-canonical generating function evaluated at a fixed particle
number $N$. This produces a probability distribution entirely
inherited from the grand-canonical setting, not from the canonical
one, so it is not surprising that (\ref{ziff}) deviates from
canonicality. It should be added that, for large $N$, the
fluctuations of $N$ in the grand-canonical ensemble become
negligible compared to $N$ itself; thus fixing $N$ does not
distort the large-scale statistics and virtually leads to the same
results that would be obtained in a fully canonical setting as
ours. More concretely, in this regime the dominant occupancy
configurations in (\ref{ziff}) coincide with those that, as in
this paper, are generated by sampling $N$ independent one-particle
with Boltzmann distribution, because both constructions
concentrate around occupancy numbers close to $N e^{-\beta
\epsilon_c}/Z$. In this perspective, we see that, compared to
\cite{ziff}, the merit of the approach developed here is that it
reconciles canonicality in the full Hilbert space and in its
symmetric subspace also for small $N$, a regime in which, due to
the non-negligible fluctuations of the number of particles in the
grand-canonical ensemble, canonicality cannot be fully captured by
(\ref{ziff}).

\section{Conclusion}
Starting from a suggestion sketched in \cite{pathria}, we have
worked out the probability distribution of the occupancy numbers
of canonical bosonic quantum systems. When canonicality is
intended asymptotically and in probability as it is in
(\ref{canprob2}), the sought distribution turns out to be
multinomial. Bosonic canonical typicality has been discussed in
the paper and upper bounds in the style of \cite{typpopescu} have
been derived. Future research would address the compatibility
between the Shannon entropy of the multinomial distribution and
the physical entropy of a canonical system, see \cite{spalvthermo}
for a preliminary analysis that shows that the Boltzmann
distribution does not guarantee maximum entropy with expected
energy constraint, hence it could be not the appropriate
categorical distribution for systems at the thermal state. Also,
the experimental results of condensation of systems made by few
particles could be analyzed by putting one against the other our
proposed multinomial distribution equipped with a properly
designed categorical distribution and the currently accepted
distribution of \cite{ziff}.


\end{document}